\documentclass[sn-basic]{sn-jnl}

\usepackage{graphicx}%
\usepackage{multirow}%
\usepackage{amsmath,amssymb,amsfonts}%
\usepackage{amsthm}%
\usepackage{mathrsfs}%
\usepackage[title]{appendix}%
\usepackage{xcolor}%
\usepackage{textcomp}%
\usepackage{manyfoot}%
\usepackage{booktabs}%
\usepackage{algorithm}%
\usepackage{algorithmicx}%
\usepackage{algpseudocode}%
\usepackage{listings}%
\usepackage{subcaption}

\begin{document}

\title[Multi-Modal Dataset Creation for Federated~Learning with DICOM Structured Reports]{Multi-Modal Dataset Creation for Federated~Learning with DICOM Structured Reports}

\author[1,2,3]{\fnm{Malte} \sur{Tölle} \orcid{0000-0002-0804-5794}}\email{malte.toelle@med.uni-heidelberg.de}
\author[1,2,3]{\fnm{Lukas} \sur{Burger}}
\author[1,2,3]{\fnm{Halvar} \sur{Kelm}}
\author[1,2,3]{\fnm{Florian} \sur{André}}
\author[1,4]{\fnm{Peter} \sur{Bannas}}
\author[5]{\fnm{Gerhard} \sur{Diller}}
\author[1,2,3]{\fnm{Norbert} \sur{Frey} \orcid{0000-0001-7611-378X}}
\author[5]{\fnm{Philipp} \sur{Garthe}}
\author[1,6]{\fnm{Stefan} \sur{Groß}}
\author[1,4,7,8,9]{\fnm{Anja} \sur{Hennemuth}}
\author[1,6]{\fnm{Lars} \sur{Kaderali}}
\author[1,7,8,9]{\fnm{Nina} \sur{Krüger}}
\author[1,10]{\fnm{Andreas} \sur{Leha}}
\author[1,11]{\fnm{Simon} \sur{Martin}}
\author[1,7]{\fnm{Alexander} \sur{Meyer} \orcid{0000-0002-6944-2478}}
\author[1,11]{\fnm{Eike} \sur{Nagel}}
\author[5]{\fnm{Stefan} \sur{Orwat}}
\author[1,12]{\fnm{Clemens} \sur{Scherer} \orcid{0000-0003-2816-6793}}
\author[1,13]{\fnm{Moritz} \sur{Seiffert}}
\author[1,4]{\fnm{Jan Moritz} \sur{Seliger}}
\author[1,6]{\fnm{Stefan} \sur{Simm} \orcid{0000-0001-9371-2709}}
\author[1,8]{\fnm{Tim} \sur{Friede}}
\author[1,14,15]{\fnm{Tim} \sur{Seidler}}
\author[1,2,3]{\fnm{Sandy} \sur{Engelhardt} \orcid{0000-0001-8816-7654}} 

\affil[1]{\small \orgname{DZHK (German Centre for Cardiovascular Research, all partner sites)}}

\affil[2]{\small \orgdiv{Department of Cardiology, Angiology and Pneumology}, \orgname{Heidelberg University Hospital}, \orgaddress{\city{Heidelberg}, \country{Germany}}}
\affil[3]{\small \orgname{Informatics for Life Institute}, \orgaddress{\city{Heidelberg}, \country{Germany}}}
\affil[4]{\small \orgdiv{Department of Diagnostic and Interventional Radiology and Nuclear Medicine}, \orgname{University Medical Center Hamburg-Eppendorf}, \orgaddress{\city{Hamburg}, \country{Germany}}}
\affil[5]{\small \orgdiv{Clinic for Cardiology III}, \orgname{University Hospital Münster}, \orgaddress{\city{Münster}, \country{Germany}}}
\affil[6]{\small \orgname{Institute of Bioinformatics}, \orgname{University Medicine Greifswald}, \orgaddress{\city{Greifswald}, \country{Germany}}}
\affil[7]{\small \orgname{Deutsches Herzzentrum der Charité (DHZC)}, \orgdiv{Institute of Computer-assisted Cardiovascular Medicine}, \orgaddress{\city{Berlin}, \country{Germany}}}
\affil[8]{\small \orgname{Charité – Universitätsmedizin Berlin}, \orgdiv{corporate member of Freie Universität Berlin and Humboldt-Universität zu Berlin}, \orgaddress{\city{Berlin}, \country{Germany}}}
\affil[9]{\small \orgname{Fraunhofer Institute for Digital Medicine MEVIS}, \orgaddress{\city{Bremen}, \country{Germany}}}
\affil[10]{\small \orgdiv{Department of Medical Statistics}, \orgname{University Medical Center Göttingen}, \orgaddress{\city{Göttingen}, \country{Germany}}}
\affil[11]{\small \orgname{Institute for Experimental and Translational Cardiovascular Imaging}, \orgname{Goethe University}, \orgaddress{\city{Frankfurt am Main}, \country{Germany}}}
\affil[12]{\small \orgdiv{Department of Medicine I}, \orgname{LMU University Hospital, LMU Munich}, \orgaddress{\city{Munich}, \country{Germany}}}
\affil[13]{\small \orgdiv{Department of Cardiology, University Heart and Vascular Center Hamburg}, \orgname{University Medical Center Hamburg-Eppendorf}, \orgaddress{\city{Hamburg}, \country{Germany}}}
\affil[14]{\small \orgdiv{Department of Cardiology}, \orgname{University Medicine Göttingen}, \orgaddress{\city{Göttingen}, \country{Germany}}}
\affil[15]{\small \orgdiv{Department of Cardiology}, \orgname{Campus Kerckhoff of the Justus-Liebig-Universität Gießen}, \orgname{Kerckhoff-Clinic}, \orgaddress{\city{Gießen}, \country{Germany}}}

\abstract{
\textbf{Purpose:} 
Federated training is often hindered by heterogeneous datasets due to divergent data storage options, inconsistent naming schemes, varied annotation procedures, and disparities in label quality.
This is particularly evident in the emerging multi-modal learning paradigms, where dataset harmonization including a uniform data representation and filtering options are of paramount importance.

\textbf{Methods:} 
DICOM structured reports enable the standardized linkage of arbitrary information beyond the imaging domain and can be used within Python deep learning pipelines with \texttt{highdicom}.
Building on this, we developed an open platform for data integration and interactive filtering capabilities that simplifies the process of assembling multi-modal datasets.

\textbf{Results:} 
In this study, we extend our prior work by showing its applicability to more and divergent data types, as well as streamlining datasets for federated training within an established consortium of eight university hospitals in Germany.
We prove its concurrent filtering ability by creating harmonized multi-modal datasets across all locations for predicting the outcome after minimally invasive heart valve replacement.
The data includes DICOM data (i.e. computed tomography images, electrocardiography scans) as well as annotations (i.e. calcification segmentations, pointsets and pacemaker dependency), and metadata (i.e. prosthesis and diagnoses).

\textbf{Conclusion:} 
Structured reports bridge the traditional gap between imaging systems and information systems. 
Utilizing the inherent DICOM reference system arbitrary data types can be queried concurrently to create meaningful cohorts for clinical studies.
The graphical interface as well as example structured report templates will be made publicly available.}


\maketitle

\section{Introduction}

The predictive performance of modern deep learning models scales with the amount of training data.
The so called foundation models are trained on a incredibly large corpus of data~\cite{oquab2023dinov2}.
A promising research direction is the incorporation of multi-modal data into the model training pipeline~\cite{chambon2022roentgen}. 
The literature on foundation models in the medical domain is scarce due to missing large scale datasets attributable to missing data unification and privacy laws.

lectronic health records (EHR) promise a unified intra- and inter-hospital data representation that improves patient care by providing a timely access to centralized patient's medical records.
Data based on EHRs is comprised of different formats including images, patient history, diagnostic reports, and other clinical domains.
Theoretically, there are many templates that can be followed to ensure interoperability between healthcare providers' data.
However, in practice many different data formats and often self-defined templates are used that hinder the collection of large scale datasets even in an intra-hospital cohort selection~\cite{noumeir06sr}.
The difficulty of data harmonization increases due to the breadth of templates used when operating across multiple locations, which is additionally impeded by privacy laws.
To realize the vision of the widespread application of deep learning models in clinical practice, standard interfaces need to be defined that enable interoperability.

One renowned method, which can circumvent privacy concerns, but is dependent on data harmonization, is  Federated learning (FL)~\cite{rieke20fl}.
FL inverts the common paradigm of central data storage by sending the model to each data owning institution, where local training is performed, before the model weights are sent back to the central instance for averaging.
However, due to the missing possibility of manually inspecting the data at each data holding institution data harmonization is of paramount importance to enable a meaningful model training.

A  possible internationally accepted standard for data harmonization is Digital Imaging and Communication in Medicine (DICOM), which is used for communication and storage of medical images and related information across a wide range of medical modalities and disciplines~\cite{roth2016dicom}.
While originating in the field of imaging, it does also allow for storage and linking of other modalities such as waveforms, text, audio, or speech in structured reports (SR).
Although hospitals around the world have established an extensive enterprise imaging infrastructure, workflows and software applications based on DICOM often rely on non-standard formats and interfaces for the storage and exchange of data annotations and computational image analysis results~\cite{bridge22highdicom}.

The aim of this work is to provide a platform for efficient data integration, matching, and cohort selection for multi-modal (federated) deep learning based on real world DICOM data.
Data integration refers to the aggregation of diverse data types into a single, coherent framework, while data matching ensures entries can be referenced and linked unambiguously on different levels of the document tree.
The contributions of this work are four-fold:
\begin{itemize}
    \item By leveraging the DICOM standard we integrate data from different sources and modalities into a coherent framework that exceeds the imaging domain by using structured reports (SR).
    \item We present a method to match and filter the data that is coming from different modalities and linked on different levels in the document tree.
    \item SRs represent a difficult domain due to their vast amount of templates. 
    We present a method for a unified extraction of the individual important information per template.
    \item By implementing an export functionality for the created cohorts, we are able to create harmonized datasets to enable federated training across multiple locations.
\end{itemize}

A previous version of this article was published at the German Conference on Medical Image Computing~\cite{toelle2024sr}.
In that version we presented the general feasibility to filter data in a single location training scenario.
In this extension we install the federated infrastructure together with our designed filter tool at all locations within our FL consortium.
Each location exports its available data from their clinical PACS, pairs it with potentially present labels, converts the data to DICOM representation, and uploads everything into the presented tool.
After performing cohort selection at each individual location, the data is exported to subsequently enable a federated training.
To enable the conversion of all data types across the consortium we generate structured reports from additional templates for a more diverse and larger set of multi-modal medical data and subsequently investigate its filterability.

\section{Material and methods}

\paragraph{Requirements}

An application that allows consistent data integration, matching, and cohort selection should fulfil four requirements.

\begin{itemize}
    \item \emph{R1 Integration}: Consistent representation of data from different formats in an unifying interface. 
    All required data shall be storable at one place in the same data format without the need to query multiple data bases.
    \item \emph{R2 Unambiguous Matching}: Different data types from the same patient must be matched and queryable together. 
    Concurrently, they must be filterable independent of other types and multiple conditions must be evaluable on the instance level.
    \item \emph{R3 Intuitive Cohort Selection}: Graphical visualization and filtering is desirable, while more complicated textual queries should also be possible.
    \item \emph{R4 Independent Platform or Extension to an Existing Platform}: Usage shall be enabled as a stand-alone application as well as an extension to existing frameworks such as e.g. Kaapana~\cite{Scherer2020JIP}.
\end{itemize}

\begin{figure}
    \centering
    \includegraphics[width=0.85\textwidth]{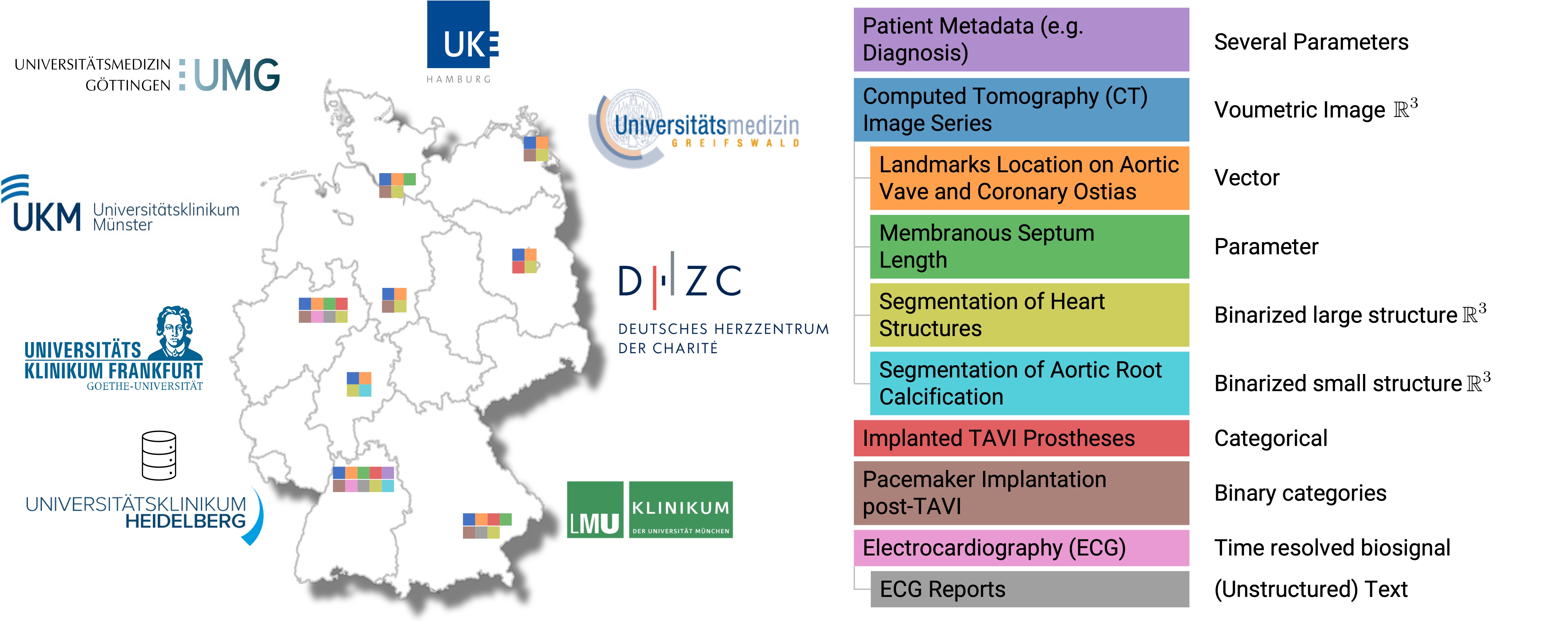}
    \caption{
        Different data types relevant for prediction of outcome after minimally invasive transcatheter aortic valve implantation (TAVI). 
        The data is heterogeneous distributed across locations in the consortium in terms of type (indicated by the color scheme) and quantity.
        Many different influence factors exist for TAVI~\cite{Genereux2012}. 
        Derived information for an original data source is indicated by the hierarchy on the right hand side.
    }
    \label{fig:consortia}

    \centering
    \includegraphics[width=0.85\textwidth]{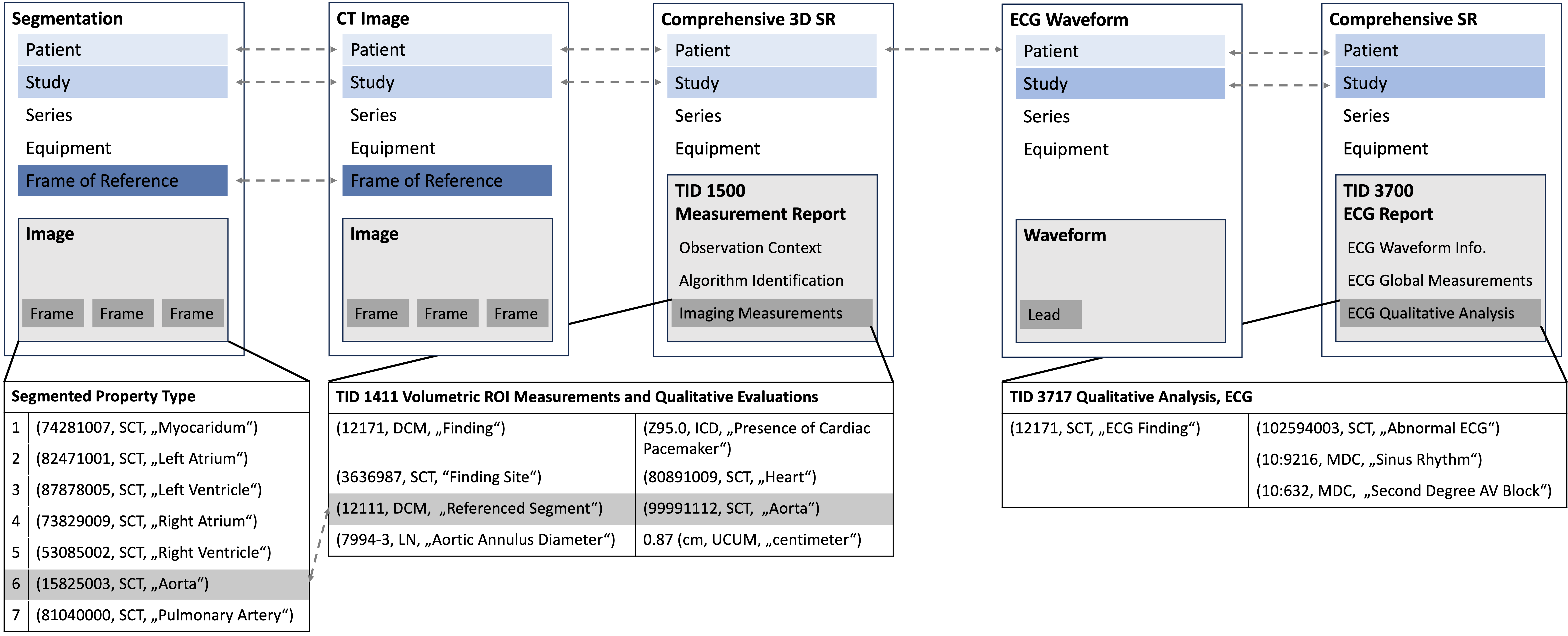}
    \caption{Relevant data and their relationship to predict pacemaker dependecy after heart valve replacement. Structured Reports can reference e.g. images, waveforms, regions, and segments. Similar to \texttt{highdicom} we also opt for the general Comprehensive (3D) SR due to its high generalizability \cite{bridge22highdicom}.}
    \label{fig:use-case}
\end{figure}

\paragraph{DICOM Annotations}

Integrating machine learning models into clinical workflows requires interoperability between existing systems, achievable through adherence to established standards.
Although the DICOM standard is mainly used within the imaging departments of hospitals it allows storage of other data types with already defined templates as well.

Structured reports represent a standardized method for the encoding, transmission, and storage of diagnostic findings.
In addition to images, they also allow the linking of other data types such as e.g. text, audio, time points, or other measurements.
Each SR must have a defined template that consists of a sequence of content items (see Fig. 6b in the appendix). 
Each content item defines a name-value pair that encodes a domain-specific property or concept.
Concept names must follow standard medical ontologies and terminologies such as the Systematized Nomenclature of Medical Clinical Terms (SNOMED CT), which ensures uniqueness and cross domain interpretability~\cite{nobel22sr}.
The value can take up to four different types: it can also be a coded concept, it might include a number, comprise one or more sets of points or other graphic data such as surfaces, or can be made up of plain text (see Fig. 5b in the appendix). 
Recently, the Python library \texttt{highdicom} was introduced, which employs an object-oriented approach towards SRs simplifying their handling within deep learning pipelines~\cite{bridge22highdicom}.
However, they have so far only implemented template identity (TID) 1500 Measurement Report as their main focus was radiology and pathology.
The implementation of other templates allows for the extension to other fields. 
Another common data type in radiology are segmentations in which one or multiple structures can be compartmentalized.
Each segment can be described by a specific label encoded by a coded concept similar to the name of a content item in SRs that ensures uniqueness.
Segments can also be referenced in a SR providing another level of entwinement.

\paragraph{Existing Platforms and Libraries}

One renowned platform that aims at facilitating the application of algorithms on real world clinical data is Kaapana\footnote{\url{https://www.kaapana.ai}}~\cite{Scherer2020JIP}.
In addition to the possibility of deploying containerized algorithms directly in the platform they also provide a graphical filter tool for cohort selection based on the imaging data in the internal picture archiving and communication system (PACS).
However, multi-modal dataset creation is not supported as only image related attributes can be filtered and links to other forms of data (diagnoses, geometric relations, etc.) are not included.

We showcase a possible integration of our developed platform by incorporation into Kaapana.
Similar to their inbuilt functionality we also opt for Opensearch as the tool of choice for enabling graphical cohort filtering\footnote{\url{https://opensearch.org/}}.
But due to its deployability in a docker container our tool can also serve as a stand alone tool, which can be linked to any PACS.
Opensarch's filtering plots are highly customizable due to the JSON based representation scheme, which also allows for very fast query times with a unified data representation.
While we believe our presented layout is a good compromise between complexity and usability, other researchers might want to include further elements or plots and subsequently use them for filtering, which Opensearch allows out of the box.

A patient can have multiple reports or segmentations, which can contain e.g. different diagnoses or labels for different studies.
Filtering these interlinked data types on the same level requires the child element to be aware of its parent due to the possible reference of multiple instances within one object.
To ensure a parent document is evaluated at the level of its child, i.e. annotation, we use Opensearch's nested objects.
With nested objects we can evaluate whether multiple conditions are true for one instance i.e. our children obtain boundaries.
As an example we might want all segmentations that have the left and right ventricle included.
Since segmentations reference a series on the whole without nested objects we would also receive all instances that have only one of both segmented.
To visualize such nested queries, one cannot use the default visualizations of Opensearch, one must rather use the declarative language Vega to create custom plots\footnote{\url{https://vega.github.io/vega/}}.

\section{Evaluation}

\paragraph{Federated Learning Consortium}

\footnotetext{\url{https://dicom.nema.org}}

Within the German Centre for Cardiovascular Diseases (DZHK)\footnote{\url{https://dzhk.de/}} and further German Hospitals we have established an FL consortium spanning across eight university clinics (see Fig.~\ref{fig:consortia}).
The use case we want to address is the outcome prediction after minimally invasive transcatheter aortic valve implantation (TAVI)~\cite{seidler22floto}.
In the literature multiple possible impact factors on outcome derived from e.g. computed tomography (CT) and electrocardiography (ECG) scans are reported, such as the length of the membranous septum or the presence of a left bundle branch block (see Fig.~\ref{fig:use-case})~\cite{Genereux2012}. 
In general, the individual ECG scans can indicate existing disturbed heart stimulus conduction, while CTs provide further information about geometric relations and calcification. 
Additionally, patient characteristics such as a history of heart disease or diabetes mellitus can yield further insides.
However, a dedicated analysis of the interactions and an automated analysis on the raw input data with deep learning across multiple institutions is missing.

\begin{figure}
    \centering
    \includegraphics[width=0.7\linewidth]{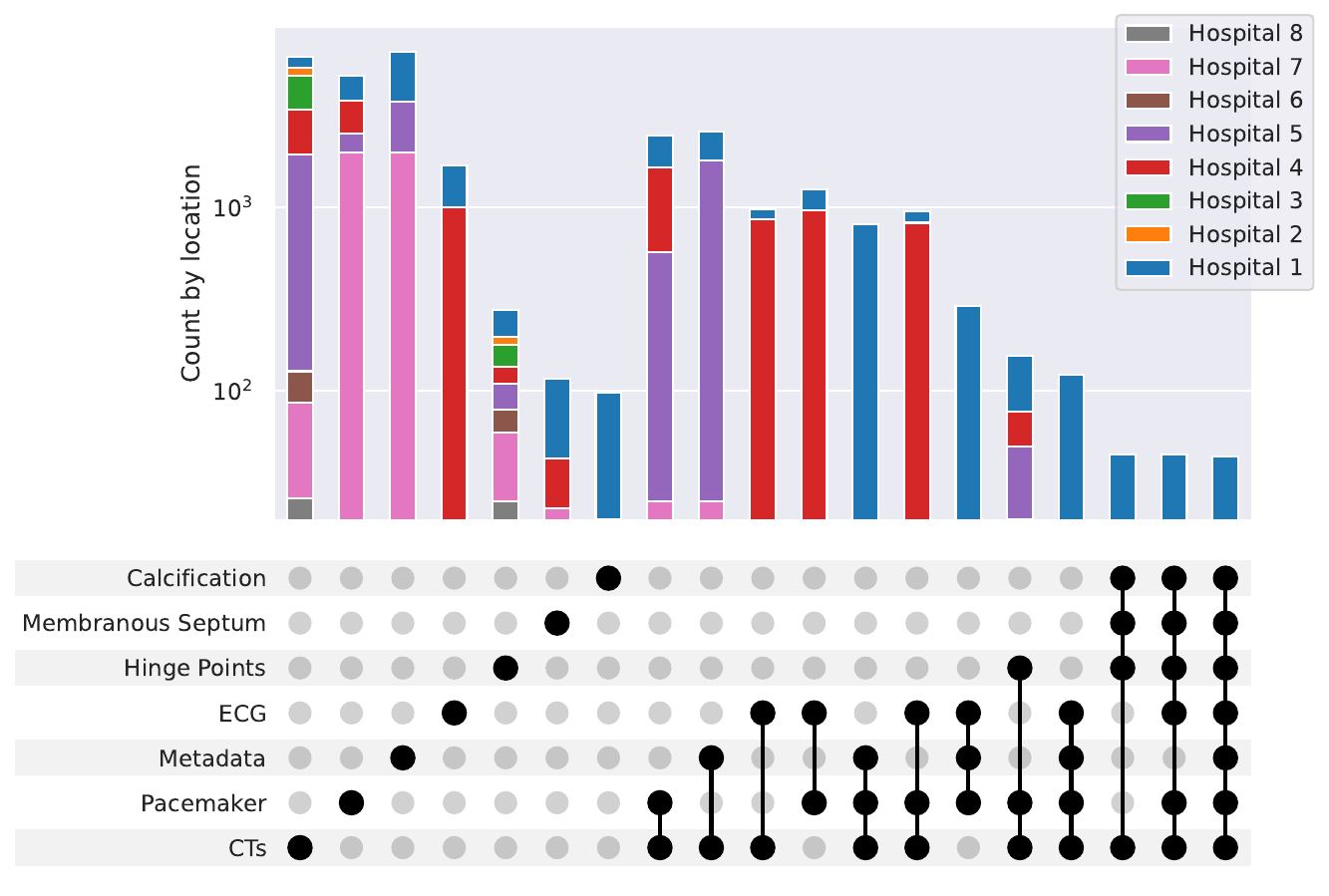}
    \caption{
    Distribution of different label subsets across locations on log-scale. 
    The lower diagram displays the composition of the subset, where we restrict the subsets based on our estimation for their usefulness for independent model training. 
    In the upper diagram the amount of samples across all locations for the particular subset of data types are visualized.
    Pacemaker refers to whether information (yes/no) of implantation is avilable.
    }
    \label{fig:label-distribution}
\end{figure}

\paragraph{Data Collection}

Every institution was required to export the relevant data they posses from their internal clinical information systems.
Even though every institution exported their CT scans from a PACS system the cross-location heterogeneity in the resulting data was large.
The textual descriptions as well as the field of view covered by the volume differed between and also within institutions.
Even worse are the data representations for ECG; it varied from extensive markup language (XML)\footnote{\url{https://www.w3.org/TR/REC-xml/}} over standard communications protocol for computer assisted electrocardiography (SCP-ECG)\footnote{\url{https://dicom.nema.org/medical/dicom/current/output/chtml/part17/sect_C.7.html}} to DICOM~\cite{roth2016dicom}.
The information of an implanted pacemaker was exported from the German procedure classification, the official classification for the encoding of operations, procedures and general medical measures\footnote{\url{https://www.bfarm.de/EN/Code-systems/Classifications/OPS-ICHI/OPS/_node.html}}.
General patient metadata was extracted from the quality insurance data for the Institute for Quality Assurance and Transparency in Healthcare\footnote{\url{https://iqtig.org/qs-verfahren/hch-komb/}}.
Prior to filtering we had to convert all data types to DICOM, to enable integration of all data types in a single data storage and matching between CTs, ECGs, and SRs on a patient level.
To verify the intended functionality of our platform we additionally uploaded five public datasets and aimed at filtering out the needed subset for our pacemaker prediction scenario at each location in our consortium~\cite{saltz2021coviddataset,martinisla2023mnm2,wasserthal2022totalsegmentator,landman2015bcv,kavur2019chaos}.

\section{Results}

Our tool allows for the simultaneous filtering of DICOM imaging and waveform modalities and annotations in the form of structured reports and segmentations (\emph{R1}).
All data types present in the consortium could be converted to SRs and subsequently be uploaded into the local PACS. 
The DICOM representation ensures matching on a patient level across time points (\emph{R2}).
In the case of segmentations we visualize all segment descriptions as well as the creator type (\emph{R3}).
The dockerized implementation allows for a stand-alone usage as well as an integration into existing frameworks (R4).

SR templates exhibit a high degree of sophistication due to their intricately nested structure and the numerous attributes that can be defined \cite{bridge22highdicom}.
To make the data filterable when uploading the SR's data into Opensearch the user must define, which measurements, geometrical data, qualitative and text attributes shall be queryable by providing the nested path (e.g. for TID 3700 ECG Report: Cardiovascular Patient History - Social History - Tobacco Smoker).
Although this might be complex at times we believe that it provides the flexibility needed for filtering such a sophisticated data structure as SRs.
The definition of the path to the wanted attributes must be defined once per template, but it can be expanded or contracted at any time (\emph{R3}).
This also allows for the dynamic creation of custom templates if needed.
An example for the nested, object-oriented representation of the SR template 3700 ECG report and 3708 waveform information in \texttt{highdicom} can be found in Fig. 6 in the appendix.

\begin{figure}
    \centering
    \includegraphics[width=.9\textwidth]{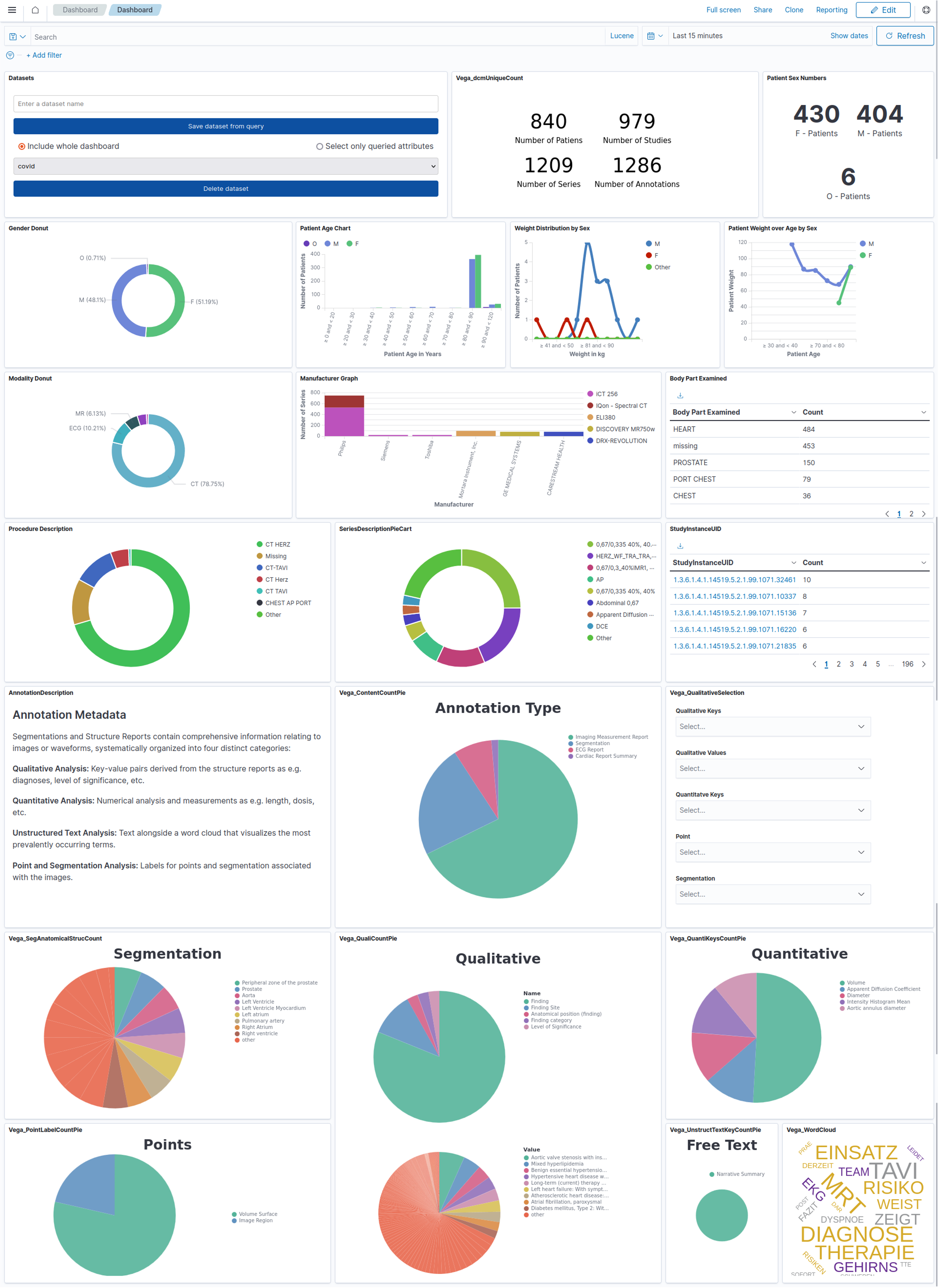}
    \caption{The created dashboard with the filterable annotation attributes. The type of annotation can be chosen. Segmentations can be queried on individual segment level. Since for qualitative items both name and value are a concept name, we can filter both. For numeric (e.g. measurements) as well as geometric (e.g. points) items we filter on the name. When the value is text we incorporate free text search. Some filter options concerning imaging modalities (e.g. manufacturer or body part examined) are similar to existing functionality in Kaapana, but under the hood our tool utilizes nested objects to enhance filterability with the given elements.}
    \label{3484-dashbaord}
\end{figure}

Our chosen method was able to extract the required information from the uploaded SR templates and subsequently visualize and modify its distribution depending on the wanted cohort.
The final dashboard for Heidelberg is shown in Fig.~\ref{3484-dashbaord}, further resulting distribution from other locations can be found in the appendix.
The interactive graphical filtering procedure is showcased in the supplemental video.

In Heidelberg we ended up with 1209 series from 840 patients from five modalities (MRI, CT, ECG, classical X-ray (CR), digital radiography (DX)), with a multitude of different annotations (1286) comprising of segmentations (21 different labels) in total.
In total we had over 20 different types of qualitative items, 5 quantitative, 2 geometric, and 1 free text.
With the help of our interface we identified 813 CT and 700 ECG scans that can generally be used for model training.
We had 78 scans with labelled hinge points and coronary arteries (HPs and CAs), 73 with membranous septum (MS), and 78 with annotated aortic root calcification.
When performing filtering at all federated locations, we obtained 6592 CT scans over all locations, 982 with CT and ECG scans across three locations, 7088 patients for which we had the implanted prosthesis information, and 5204 patients for which we had the label of whether a pacemaker was implanted or not post-TAVI.
The quantities within different subsets of data are shown in Fig.~\ref{fig:label-distribution}.
After exporting all datasets in a unified manner to be readable by a federated learning pipeline, we can progress with training on each individual subset or the intersection of two or more.
The training results will be published elsewhere.

In general the amount of pacemaker dependent patients after TAVI is low at each individual institution.
Thus, we have a very unbalanced label distribution if trained only at one location.
By employing the federated data collection we were able to obtain more cases and thus a more favorable balancing of the data the overall distribution of pacemaker labels.

\section{Discussion}

Despite their complexity, SRs provide an useful tool for unifying diagnostic reports for physicians, model training, and results when utilized appropriately.
The DICOM standard unambiguously matches patient data samples, but large cohort creation that includes patients based on specific attributes is difficult from a PACS interface.
Our solution for determining on which attributes to filter the annotations opts for high flexibility as it is applicable to all SR templates by concurrent conformance with other DICOM data objects such as images or waveforms.
We believe our method is a step towards the concurrent filtering and cohort selection of DICOM data and their annotations.
The platform is openly available and can be used in other projects.

SRs still have more inherent complexity than we have covered here, which is the main reason they are not adopted widely in clinical practice yet, despite their obvious usefulness.
The adoption of the object-oriented approach in \texttt{highdicom} for defining and using SR templates in Python is a step towards their more widespread usage~\cite{Fedorov2023sr2,Krishnaswamy2024sr2}.
Still, the definition of the required attributes for creating a SR is tedious and requires manual effort.
In future work, we also aim at storing the model output in a consistent manner similar to the homogenized training data.

\section*{Declarations}

\bmhead{Conflict of interest}
Norbert Frey reports speaker honoraria, presentations or advisory board consultations from AstraZeneca, Bayer AG, Boehringer Ingelheim, Novartis, Pfizer, Daiichi Sankyo Deutschland.
Tim Seidler reports research, educational, or travel grants and honoraria for lectures or advisory board consultations from Abbott Vascular, AstraZeneca, BoehringerIngelheim, Bristol Myers Squibb, Corvia, Cytokinetics, Edwards Life Sciences, Medtronic, Myocardia, Novartis, Pfizer, Teleflex.
Alexander Meyer reports consulting or lecturing fees from Medtronic, Bayer, Pfizer.
Clemens Scherer reports speaker honorarium from AstraZeneca.
None are related to the content of the manuscript
The other authors declare no conflicts of interest.

\bmhead{Funding}
The project was funded by the DZHK, the Klaus Tschira Foundation within the Informatics for Life framework, and the BMBF-SWAG Project 01KD2215D.

\bmhead{Ethical approval}
All procedures performed in studies involving human participants were in accordance with the ethical standards of the institutional and/or national research committee and with the 1964 Helsinki declaration and its later amendments or comparable ethical standards.
This article does not contain any studies with animals performed by any of the authors.
Ethical approval was waived by the local Ethics Committees of Heidelberg (S-475/2021), Göttingen (11/6/21), Hamburg (2021-200262-BO-bet), Munich (21-0497), Münster (2021-487-b-S), and Frankfurt (2021-366\_1) in view of the retrospective nature of the study and all the procedures being performed were part of the routine care.
In Berlin, a multi-centric study must not explicitly be confirmed when another institutional ethics board waived approval.

\bmhead{Informed consent}
Informed consent was not obtained due to the retrospective nature of the study.

\bibliography{sn-bibliography}

\end{document}